\documentclass[12pt]{article}
\usepackage{latexsym}
\usepackage{epsf}
\usepackage{hyperref}

\def\bibi{\bibitem}

\def\a{\alpha}
\def\b{\beta}

\def\d{\delta}
\def\g{\gamma}

\def\m{\mu}
\def\n{\nu}

\def\p{\pi}                     
\def\s{\sigma}                  
\def\t{\tau}

\def\x{\xi}

\def\L{\Lambda}

\def\cd{{\cal D}}

\def\cf{{\cal F}}
\def\cg{{\cal G}}
\def\ch{{\cal H}}   

\def\cl{{\cal L}}

\def\cbo{{\,\raise-.15ex\Sc [\,}}                       

\def\ddt#1{{\buildrel {\hbox{\LARGE .\kern-2pt.}} \over {#1}}}

\def\secteq#1{ \setcounter{equation}{0}
               \renewcommand{\theequation}{#1.\arabic{equation}} }

\def\ie{\mbox{\it i.e.} }

\def\tr{{\rm tr}\,}

\def\half{{1\over 2}}

\def\bibi{\bibitem}       

\thicklines                         
\def\Cbar{{\overline{C}}}

\def\tg{\tilde{g}}
\def\tL{\tilde{\L}}
\def\tb{\tilde{b}}
\def\tc{\tilde{c}}
\def\td{\tilde{d}}
\def\tit{\tilde{t}}
\def\hb{\hat{b}}

\def\ie{{\it i.e.}}

\def\OUT#1{}

\def\ssec#1#2{
  \vskip 3ex
  \noindent {\bf #1 #2}\\
  \vskip -1ex
  \hskip -4ex}

\begin{document}
\hyphenation{fer-mio-nic per-tur-ba-tive pa-ra-me-tri-za-tion
pa-ra-me-tri-zed a-nom-al-ous}

\renewcommand{\thefootnote}{$*$}

\rightline{ROME1-1418/2005}
\begin{center}
\vspace{10mm}
{\large\bf Running couplings in equivariantly gauge-fixed\\[4mm]
$SU(N)$ Yang--Mills theories}
\\[12mm]
Maarten Golterman$^a$\footnote{Permanent address: Department of Physics and Astronomy,
San Francisco State University,
San Francisco, CA 94132, USA}\ and \ Yigal Shamir$^b$
\\[8mm]
{\small\it
$^a$Dipartimento di Fisica,
Universit\`a di Roma  ``La Sapienza"\\
INFN, Sezione di Roma  ``La Sapienza"\\
P.le A. Moro 2, I-00185 Roma, Italy}\\
{\tt maarten@stars.sfsu.edu}
\\[5mm]
{\small\it $^b$School of Physics and Astronomy\\
Raymond and Beverly Sackler Faculty of Exact Sciences\\
Tel-Aviv University, Ramat~Aviv,~69978~ISRAEL}\\
{\tt shamir@post.tau.ac.il}
\\[10mm]
{ABSTRACT}
\\[2mm]
\end{center}

\begin{quotation}
In equivariantly gauge-fixed $SU(N)$ Yang--Mills theories, the gauge symmetry is
only partially fixed, leaving a subgroup $H\subset SU(N)$ unfixed.
Such theories avoid Neuberger's nogo theorem if the subgroup $H$
contains at least the Cartan subgroup $U(1)^{N-1}$, and they
are thus non-perturbatively well defined if regulated on a finite lattice.
We calculate the one-loop beta function for the coupling $\tg^2=\xi g^2$,
where $g$ is the
gauge coupling and $\xi$ is the gauge parameter, for
a class of subgroups including
the cases that $H=U(1)^{N-1}$ or $H=SU(M)\times SU(N-M)\times U(1)$.  The coupling $\tg$
represents the strength of the interaction of the gauge degrees of freedom
associated with the coset $SU(N)/H$.  We find that $\tg$, like $g$, is
asymptotically free.
We solve the renormalization-group equations for the running of the couplings
$g$ and $\tg$, and find that
dimensional transmutation takes place also for the coupling $\tg$,
generating a scale $\tL$ which can be larger than or equal to the scale
$\L$ associated with the gauge coupling $g$, but not smaller.  We
speculate on the possible implications of these results.
\end{quotation}

\renewcommand{\thefootnote}{\arabic{footnote}}
\setcounter{footnote}{0}

\newpage
\vspace{5ex}
\noindent {\large\bf 1.~Introduction}
\secteq{1}
\vspace{3ex}

In a recent paper \cite{gsnonab} we constructed equivariantly gauge-fixed $SU(N)$
Yang--Mills (YM) theories on the lattice, generalizing earlier work for the gauge group
$SU(2)$ \cite{schaden}.   The aim was to gauge fix as much of the $SU(N)$ gauge
symmetry as possible while avoiding a nogo theorem stating that the standard
BRST procedure breaks down non-perturbatively on the lattice in the compact
formulation \cite{hn}.  It turns out that it is possible to gauge fix $SU(N)$ down to
the maximal abelian subgroup $H=U(1)^{N-1}$ while avoiding the theorem.

There are other ways to gauge fix YM theories non-perturbatively,
but usually the gauge-fixing procedure is extraneous to the path integral
(for a review, see ref.~\cite{gfreview}).  Our
aim was to maintain explicit locality as well as a version of BRST invariance.
At the price of leaving (at least) the Cartan subgroup unfixed, equivariantly gauge-fixed
theories fulfill this aim: Their complete
definition consists of a path integral with a local action which, in turn, is
invariant under an equivariant BRST (eBRST) symmetry.   They are rigorously
well-defined on a finite lattice, and one can prove that correlation functions
of gauge-invariant operators in the un-fixed theory are exactly the same as
those in the equivariantly gauge-fixed theory.  The eBRST symmetry
is still nilpotent in the following sense:  The eBRST algebra closes on the
subgroup $H$, which is left unfixed.  This implies that on $H$-invariant operators eBRST
is still nilpotent, and thus the usual consequences of nilpotency still apply.

Equivariantly gauge-fixed $SU(N)$ YM theories, in which a subgroup $H$ containing
at least the Cartan subgroup of $SU(N)$ remains unfixed, are thus non-perturbatively
well defined, local gauge theories, possessing an exact BRST-like invariance.
Of course, if one would like to completely gauge fix the theory non-perturbatively, the remaining subgroup
$H$ would still need to be fixed.   At present, it is not known how to do this such that
both the nogo theorem is avoided and a form of BRST symmetry is preserved.\footnote{
For a way out for purely abelian theories, see ref.~\cite{testa}.  It is not known how to generalize
this trick to our case, where the abelian degrees of freedom are only part of the
compact gauge field represented by the link variables.}   In contrast, in
perturbation theory, complete gauge fixing is always possible, and, in fact,
necessary.  In the equivariant case, we only need to further fix the subgroup $H$ in order
to set up perturbation theory.   This leads to modifications of
the equivariant Slavnov--Taylor identities, but these can be taken into account.
Unitarity and renormalizability to all orders are still guaranteed (for the case that $H$ is the Cartan
subgroup, this was investigated in detail in ref.~\cite{gsnonab}).

In this paper, we want to start a dynamical investigation of the phase diagram of
equivariantly gauge-fixed YM theories (without any fermions or other matter fields).
We would like to know whether any  phases exist other than the one containing
the continuum limit corresponding to the standard YM universality class.
In particular, we would like to know whether any phases exist in which
eBRST invariance and/or  any of the  other global symmetries  exhibit
spontaneous symmetry breaking (SSB).  We emphasize that this is a
completely non-perturbative issue: equivariantly gauge-fixed YM theories
reproduce standard perturbation theory \cite{gsnonab} (including perturbative
unitarity), and the perturbative expressions for any gauge-invariant quantity
are thus equal to those obtained in YM theory in any standard perturbative
gauge.

Let us first assess the fact that also non-perturbatively the standard YM universality class can be
obtained from its equivariantly gauge-fixed cousin.
As already stated above, gauge-invariant
correlation functions remain unchanged after equivariant gauge fixing on the
lattice in a finite volume.  In this situation, it is therefore clear that whatever
happens in the gauge-fixing (``unphysical") sector of the theory cannot
``spill over" to the physical sector: the physics should be that of the unfixed theory.
If this is true in any finite volume, it remains true in the infinite-volume limit,
and we therefore recover the standard YM universality class.

This raises the question whether SSB could occur at all in such a way as to
modify not only the gauge-variant, but also gauge-invariant quantities.
As is well-known, in order to investigate possible SSB in any finite
volume,
an external field needs to be turned on in order to ``seed" the SSB, and this external
field is
only taken away after the infinite-volume limit is taken.  Since this external
field will break eBSRT symmetry explicitly, it is not {\it a priori} clear what the dynamical
effects of SSB will be, {\it even} for the gauge invariant sector of the theory.
One anticipates that the parameters of the theory (the gauge coupling $g$ and the
gauge-fixing parameter $\x$) can be chosen such that the usual YM universality
class can be recovered.   However, whether this will happen or not is a
question for the dynamics of the theory:  there is no obvious reason that the
vanishing-external-field and infinite-volume limits will commute everywhere in the
phase diagram.  A separate question is whether any phase with SSB might
or might not contain any critical point corresponding to an interesting continuum limit.

We take as our primary point of view that it is
interesting to understand the phase diagram of an equivariantly gauge-fixed
YM theory in its own right.  This is
corroborated by our main result, that also the coupling constant
governing the longitudinal sector of our theory is asymptotically free.
Asymptotically-free couplings are rare, and suggest strong-coupling
dynamics, like bound states and/or SSB,  at low energies.

Another motivation for
our interest in the phase diagram of equivariantly gauge-fixed YM theories
comes from our interest in the
non-perturbative study of chiral gauge theories  with the help of the lattice  \cite{mg2000}.
Lattice-regulated chiral fermions break the gauge
(and, hence, any BRST-type) symmetry explicitly \cite{rome,u1,gsnonab},
and it is therefore important to explore the possible dynamics of
equivariantly gauge-fixed theories in a wider context than that in which the
infinite-volume limit is taken while maintaining eBRST symmetry rigorously
in the regularized theory.

Let us consider this in a little more detail.  On the lattice, with chiral fermions,
eBRST is broken by the regulator, \ie\ by irrelevant terms (defined as terms in the
lagrangian which vanish in the classical continuum limit).  Because of this,
counter terms are required \cite{rome,u1,gsnonab}, whose coefficients will
need to be tuned in order to recover a quantum continuum limit.  A number of
these counter terms break eBRST, and their coefficients can thus be considered
external fields in the sense discussed above.  The same reasoning thus applies:
It is a dynamical question whether SSB of eBRST (and/or any other global symmetries)
takes place, and what the consequences of this might be for gauge-invariant
quantities.

Returning to the pure YM case, the number of gauge couplings in the gauge-fixing
sector depends on the subgroup.
If we choose
$H=U(1)^{N-1}$ or $H=SU(M)\times SU(N-M)\times U(1)$ (which is a maximal
subgroup of $SU(N)$), the theory has just two couplings, which  we will take to be the square of the
gauge coupling, $g^2$, and the product of the gauge-fixing parameter $\x$ and
$g^2$, denoted by $\tg^2=\x g^2$.  We will refer to this latter
coupling as the ``longitudinal" coupling, since it determines the dynamics
of the ``reduced" theory \cite{gsnonab} in which the gauge field is restricted
to be pure gauge.
As we already said, in order to study the phase
diagram a non-perturbative study is required.  Such a study is
outside the scope of this paper.  But, it is imperative, and interesting, to begin with an
investigation of the scaling behavior of the theory in the weak-coupling
regime (small $g^2$ and $\tg^2$).   In this paper, we calculate the
beta function for $\tg^2$ to one loop (the one-loop beta function for $g^2$ is known
of course), and solve the resulting renormalization-group (RG) equations.
It turns out that also the coupling $\tg^2$ is asymptotically free, and, just
like $g^2$, generates its own associated infrared scale.

For the calculation of one-loop beta functions, we may use
the continuum theory because they are independent
of the regulator chosen, and thus most easily calculated using dimensional
regularization.   Therefore, in Sec.~2, we summarize only the continuum form
of the theory (for the lattice version, we refer to ref.~\cite{gsnonab}).  In Sec.~3, we present
the one-loop beta functions,
and we solve the resulting RG equations in Sec.~4.
In Sec.~5 we speculate on the interpretation of our results.  We conclude in
Sec.~6.  In Appendix A we summarize some group-theoretical details,
and discuss a wider class of subgroups.

\vspace{5ex}
\noindent {\large\bf 2.~Lagrangian for equivariantly gauge-fixed $SU(N)$ theory}
\secteq{2}
\vspace{3ex}

We start with the lagrangian for an equivariantly gauge-fixed $SU(N)$
YM theory.     The YM lagrangian for a gauge field $V_\m=V_\m^a T^a$
with gauge coupling $g$ can be written
as\footnote{Throughout this paper we work in euclidean space.}
\begin{eqnarray}
\label{YMlag}
\cl_{YM}&=&\frac{1}{2g^2}\;\tr(F_{\m\n}^2)\ ,\\
iF_{\m\n}&=&[D_\m(V),D_\n(V)]\ ,\nonumber\\
D_\m(V)&=&\partial_\m+iV_\m\ .\nonumber
\end{eqnarray}
We take the hermitian generators $T^a$ of the group to satisfy $\tr(T^aT^b)=\half\d_{ab}$
and define the structure constants $f_{abc}$ through $[T^a,T^b]=if_{abc}T^c$.  The
structure constants are completely anti-symmetric in all three indices.

In order to define the gauge-fixing and ghost parts of the lagrangian, we
divide the generators into two sets.  One set, $\{T^i\}$, generates the subgroup $H$,
which is left unfixed,
while the remaining generators, $\{T^\a\}$, span the coset $G/H$, with $G=SU(N)$.  We will use
indices $i,j,k,\dots$ to indicate $H$ generators, $\a,\b,\g,\dots$ for $G/H$ generators,
and $a,b,c,\dots$ for generators of the full group, $G$.  We have that
\begin{equation}
\label{frelation}
f_{i\a j}=-f_{\a ij}=-f_{ij\a}=0\ ,
\end{equation}
because the product of two elements of $H$ is again in $H$.

Correspondingly, we split the gauge field into two parts,
\begin{eqnarray}
\label{split}
V_\m&=&A_\m+W_\m\ ,\\
A_\m&=&A_\m^iT^i\ ,\ \ \ \ \  W_\m=W_\m^\a T^\a\ ,\nonumber
\end{eqnarray}
and we introduce coset-valued ghost and anti-ghost fields
\begin{equation}
\label{ghosts}
C=C^\a T^\a\ ,\ \ \ \ \ \Cbar=\Cbar^\a T^\a\ .
\end{equation}
The gauge-fixing condition $\cf(V)$ has to be covariant under local
$H$ transformations.  The simplest generalization of the usual Lorenz
gauge is obtained by taking the derivative to be $H$-covariant, and
we thus choose
\begin{equation}
\label{gaugecondition}
\cf(V)=\cd_\m(A)W_\m\equiv\partial_\m W_\m+i[A_\m,W_\m]\ .
\end{equation}
The gauge-fixing lagrangian is then given by \cite{gsnonab}
\begin{eqnarray}
\label{lgfonshell}
\cl_{gf}&=&{1\over\tg^2}\ \tr\left(\cd_\mu(A)W_\mu\right)^2\\
&&-2\ \tr\left(\Cbar\cd_\mu(A)\cd_\mu(A)C\right)
+2\ \tr\left([W_\mu,\Cbar]_\ch[W_\mu,C]_\ch\right)\nonumber\\
&&+i\ \tr\left((\cd_\mu(A)\Cbar)[W_\mu,C]+
[W_\mu,\Cbar](\cd_\mu(A)C)\right)\nonumber\\
&&+\tg^2\Biggl(\tr\left(\left(\Cbar^2\right)_{\cg/\ch}
\left(C^2\right)_{\cg/\ch}\right)
+{3\over 4}\ \tr\left(\{\Cbar,C\}_{\cg/\ch}\right)^2
+\tr\left(\left(\{\Cbar,C\}_\ch\right)^2\right)\Biggr)\,.\nonumber
\end{eqnarray}
Here the subscripts $\ch$ and $\cg/\ch$  indicate projections
onto the linear space spanned by the generators of $H$ and $G/H$,
respectively.
The full lagrangian,
obtained by adding Eq.~(\ref{YMlag}) and Eq.~(\ref{lgfonshell}), is invariant
under (on-shell) equivariant BRST (eBRST) transformations,
local $H$ transformations, and various other global symmetries.
A complete discussion of these symmetries, the off-shell form, and
the construction of this lagrangian can be found in ref.~\cite{gsnonab}.
We note that if $V_\m$ is chosen to be pure gauge, so that $F_{\m\n}$
vanishes, indeed the ``longitudinal" coupling $\tg^2$ governs the
dynamics of the theory, since it is the only coupling which occurs
in Eq.~(\ref{lgfonshell}).

In order to derive the Feynman rules for this theory, several steps
are still necessary.  First, in order to develop a perturbative expansion
in $g$, we rescale the gauge fields, $A_\m\to gA_\m$, $W_\m\to gW_\m$.
Doing this, the coupling in front of the quadratic term for $W_\m$ in
Eq.~(\ref{lgfonshell}) becomes
\begin{equation}
\label{xi}
\frac{1}{\xi}\equiv \frac{g^2}{\tg^2}\ ,
\end{equation}
in which we identify $\x$ as the usual gauge-fixing parameter.   The $W$
propagator in momentum space is given by
\begin{equation}
\label{Wprop}
\frac{1}{p^2}\left(\d_{\m\n}-\frac{p_\m p_\n}{p^2}\right)+\xi\;\frac{p_\m p_\n}{(p^2)^2}\ .
\end{equation}
Note that the
coupling $\tg^2$ still multiplies the four-ghost terms after the
rescaling of the gauge fields.  They
will play no role in this paper.\footnote{In perturbation theory, these
couplings play a role in keeping the ghosts massless \cite{gz}.}
Then, in order to define the $A_\m$ propagator, also the remaining
gauge symmetry $H$ needs to be fixed.  We do this
by choosing simply a straightforward Lorenz gauge for the field $A_\m$,
\ie\ we add
\begin{equation}
\label{Hfix}
\cl_{H}={1\over\a}\ \tr\left(\partial_\m A_\mu\right)^2
+\mbox{$H$-ghost terms}
\end{equation}
(after the redefinition $A_\m\to gA_\m$), where
$\a$ is another gauge-fixing parameter.  If $H$ is
non-abelian, an $H$-ghost sector needs to be added;
in Eq.~(\ref{Hfix}) the $H$-ghost sector is the standard one.
However, for our one-loop calculation we will not need this sector.
We emphasize that
$H$ gauge fixing is only done in order to set up perturbation theory;
the non-perturbative theory we are interested in is the (lattice version
of) the sum of Eq.~(\ref{YMlag}) and Eq.~(\ref{lgfonshell}), without Eq.~(\ref{Hfix}).
We also recall that the construction of equivariantly gauge-fixed
$SU(N)$ YM theories goes through on the lattice, with all symmetries preserved
except space-time symmetries (which are reduced to the usual
lattice space-time symmetries).  For details, we refer to
 ref.~\cite{gsnonab}.

\vspace{5ex}
\noindent {\large\bf 3.~One-loop beta functions }
\secteq{3}
\vspace{3ex}

In order to obtain the one-loop beta function for $\tg^2$, we need to calculate
the scale dependence of the gauge parameter $\xi$.  This can then be
combined with the known beta function for the gauge coupling $g^2$,
\begin{equation}
\label{bgsq}
\frac{dg^2}{dt}=-\frac{1}{16\pi^2}\; b g^4\ ,\ \ \ \ \ b=\frac{22}{3}\;N
\end{equation}
in order to obtain the beta function for $\tg^2$.
Here $t=\log(\mu/\L)$,
with $\m$ the renormalization scale, and with $\L$ the integration constant.

The scale dependence of $\x$ can be found from the ratio of the transverse
and longitudinal parts of the vacuum polarization for the coset gauge field $W$.
The following types of diagrams contribute to this vacuum polarization.
First, there are diagrams with
the usual vertices from Eq.~(\ref{YMlag}), which lead to contributions
proportional to
\begin{equation}
\label{YMvertices}
\sum_{cd} f_{\a cd}f_{\b cd}=C_2\d_{\a\b}=N\d_{\a\b}\ ,
\end{equation}
where $C_2$ is the quadratic Casimir for the adjoint representation,
equal to $N$ for $SU(N)$.
Then, there are
vertices coupling the $W_\mu$ to the $H$ gauge field $A_\mu$ coming from the first
term in Eq.~(\ref{lgfonshell}).  Therefore, there are extra contributions involving only the
structure constants $f_{\a j\g}$ (recall that $f_{ij\g}=0$).  The combination
that appears in the $W$ vacuum polarization is
\begin{equation}
\label{H}
\sum_{i\g}f_{\a i\g}f_{\b i\g }=H_\a\d_{\a\b}\ ,
\end{equation}
which defines the constants $H_\a$.
Here, we will restrict ourselves to cases where $C_H\equiv H_\a$ is independent of $\a$.%
\footnote{In these cases, one may verify that there is indeed only one longitudinal
coupling $\tg^2$, meaning that each component of $W_\m$ has the same
transverse and longitudinal renormalization constants.  For $H=U(1)^{N-1}$
this follows from invariance under the
``skewed" permutation group discussed in ref.~\cite{gsnonab}; for
$H=SU(M)\times SU(N-M)\times U(1)$ it follows from invariance under $H$.
For some other subgroups, we refer to Appendix A.}
For the maximal abelian subgroup $H=U(1)^{N-1}$ one finds that this is the
case, with $C_H=1$.  Another example where $H_\a$ turns out to be independent
of $\a$ is the (maximal) subgroup $H=SU(M)\times SU(N-M)\times U(1)$,
for which $C_H=N/2$.
An additional contribution involving $C_H$ comes from the third term
in Eq.~(\ref{lgfonshell}), because of the $\ch$ projections on the commutators;
however, at one loop this is a tadpole that does not contribute to
wave-function renormalization constants.
Finally, there are ghost-loop contributions coming from the fourth and fifth terms (third line)
in Eq.~(\ref{lgfonshell}), involving only the structure constants $f_{\a\g\d}$, in the
combination
\begin{equation}
\label{G}
\sum_{\g\d}f_{\a\g\d}f_{\b\g\d}=G_\a\d_{\a\b}\ ,
\end{equation}
which defines $G_\a$.  The constants $H_\a$ and $G_\a$ are related, because
\begin{equation}
\label{GH}
C_2\;\d_{\a\b}=\sum_{cd}f_{\a cd}f_{\b cd}=
2\sum_{i\g}f_{\a i\g}f_{\b i\g }+\sum_{\g\d}f_{\a\g\d}f_{\b\g\d}
=(2H_\a+G_\a)\d_{\a\b}
\end{equation}
(we used again that $f_{\a ij}=0$, {\it c.f.} Eq.~(\ref{frelation})).
It follows that $H_\a\le N/2$.
If $C_H=H_\a$ is independent of $\a$, so is
$C_G\equiv G_\a$, and $C_2=2C_H+C_G$ with $C_H\le N/2$.  For $H=U(1)^{N-1}$
we thus have that
$C_G=N-2$; while $C_G=0$ for $H=SU(M)\times SU(N-M)\times U(1)$
(all $f_{\a\b\g}=0$).
 For some group-theoretical
details, see Appendix A, where we prove Eq.~(\ref{H}) (and thus Eq.~(\ref{G})) and calculate $H_\a$
for some examples.

Since the $A$ field does appear on
the loops contributing to the $W$ vacuum polarization, we need the $A$
propagator.  This is why we fixed also the $H$ gauge, as already discussed in the
previous section.  If $H$ is non-abelian, $H$ ghosts appear, but they
do not couple to the $W$ fields, and therefore do not contribute to the
$W$ vacuum polarization at one loop.  Also, we expect the one-loop
beta function for $\tg^2$ to be independent of the second gauge parameter
$\a$, as is indeed borne out by explicit calculation.  The one-loop beta function
for $\tg^2$ is thus independent of the gauge used to fix the
remaining gauge group $H$.

The actual calculation of the $W$ vacuum polarization is
straightforward.  Using dimensional regularization and minimal subtraction,
we find for the one-loop transverse and longitudinal wave-function
renormalization constants
\begin{eqnarray}
\label{Zs}
Z_{trans}&=&1+\frac{g^2}{16 \pi^2 \epsilon}\;
               \left( \left(\frac{13}{3} - \xi\right)C_G + \left(\frac{17}{3} -\xi - 2\alpha\right)C_H \right)\ ,\\
Z_{long}&=&1+\frac{g^2}{16 \pi^2 \epsilon}\;
              \left(-\half\;\xi \;C_G + \left(\frac{6}{\xi} + 3-2\alpha + \xi\right)C_H \right)\ .\nonumber
\end{eqnarray}
The renormalization constant $Z_\x$ relating bare and renormalized couplings
by $\x_0=Z_\x\;\x$ is equal to the ratio of $Z_{trans}$ and $Z_{long}$, and combining
this with Eq.~(\ref{bgsq})
we obtain the $\tg^2$ beta function:\footnote{For
$N=2$ our result agrees with the result obtained in ref.~\cite{schaden2}.
For more general subgroups than the ones considered here, $C_H$ and $C_G$ in Eq.~(\ref{btgsq})
get replaced by $H_\a$ and $G_\a$, so that each new value of
$H_\a$ (and thus $G_\a$) corresponds to a new running coupling in the theory.
}
\begin{eqnarray}
\label{btgsq}
\frac{d\tg^2}{dt}&=&-\frac{1}{16\pi^2}\left(\tb g^4+\tc g^2\tg^2+\td\tg^4\right)\ ,\\
\tb&=&6C_H \ , \nonumber\\
\tc&=&b-\frac{13}{3}\;C_G-\frac{8}{3}\;C_H=
12C_H+3C_G=3N+6C_H   \ , \nonumber\\
\td&=&2C_H+\half\; C_G=\half N+C_H \ . \nonumber
\end{eqnarray}
{}From this result we conclude that the coupling $\tg^2$ is asymptotically
free, just like $g^2$, because all three coefficients $\tb$, $\tc$ and $\td$ are
positive.  Unlike Eq.~(\ref{bgsq}) though, the differential equation~(\ref{btgsq}) is inhomogeneous, and,
even if $\tg^2$ is chosen vanishingly small at some scale,
the non-zero value of $g^2$ generates a non-vanishing $\tg^2$ below that scale.  The
origin of this is the term proportional to $1/\x$ in $Z_{long}$, {\it cf.} Eq.~(\ref{Zs}).

It turns out that the RG
equation~(\ref{btgsq}) can be solved analytically.  This will be the
topic of the next section.

\vspace{5ex}
\noindent {\large\bf 4.~Running couplings and $\L$ parameters }
\secteq{4}
\vspace{3ex}

We begin by rescaling
\begin{equation}
\label{rescale}
t\to 16\pi^2 t=\log(\m/\L)\ ,
\end{equation}
in order to get rid of the factors $1/(16\pi^2)$ multiplying both beta functions.
The solution of Eq.~(\ref{bgsq}) can then be written as
\begin{equation}
\label{gsqsol}
g^2=\frac{1}{bt}\ ,
\end{equation}
where we absorbed the integration constant into the scale $\L$.  In other words,
we define $\L$ as the scale where $g^2$ diverges.
We substitute this
solution into Eq.~(\ref{btgsq}), and introduce a new variable
\begin{equation}
\label{x}
x=\tg^2+\frac{\tc}{2\td}\;g^2=\tg^2+3g^2\ .
\end{equation}
Equation~~(\ref{btgsq}) then takes the form
\begin{equation}
\label{ricatti}
\frac{dx}{dt}=-\frac{\hb}{b^2}\,\frac{1}{t^2}-\td\,x^2\ ,
\end{equation}
in which
\begin{equation}
\label{bhat}
\hb=\tb-\frac{\tc^2}{4\td}+\frac{b\tc}{2\td}=\frac{35}{2}\;N-3C_H>0\ .
\end{equation}
The positivity of $\hb$ follows because $C_H\le N/2$.
Equation~~(\ref{ricatti}) has the form of a Ricatti equation \cite{gr}, and can be exactly
solved; its solution is ($t>0$)
\begin{equation}
\label{sol}
x(t)=\frac{1}{\td t}\;\frac{p\a_+t^{\a_+}+q\a_-t^{\a_-}}{pt^{\a_+}+qt^{\a_-}}\ ,
\end{equation}
where $p$ and $q$ are two arbitrary constants (not both equal to zero), and
$\a_\pm$ are the solutions of $\a(\a-1)+\hb\td/b^2=0$,
\begin{equation}
\label{alphas}
\a_\pm=\frac{1}{2}\left(1\pm\sqrt{1-4\;\frac{\hb\td}{b^2}}\;\right)
=\frac{1}{2}\left(1\pm\sqrt{\frac{\frac{169}{4}N^2-144NC_H+27C_H^2}{121N^2}}\;\right)\ .
\end{equation}
Obviously, we are looking for real solutions, and this is what we find if $p$ and $q$
are chosen real and
the radical in Eq.~(\ref{alphas}) is non-negative.  If the radical  is negative, the solution of
Eq.~(\ref{ricatti}) may be written differently.  Introducing
\begin{equation}
\label{gamma}
\g=\sqrt{\frac{\hb\td}{b^2}-\frac{1}{4}}\ ,
\end{equation}
the solution is
\begin{equation}
\label{sol2}
x(t)=\frac{1}{2\td t}\left(1+\frac{2\g}{\tan{(\g\log(t)-\phi)}}\right)\ ,
\end{equation}
with $\phi$ an integration constant.

The two types of solutions, Eqs.~(\ref{sol}) and~(\ref{sol2}) are rather different mathematically.
Taking $H$ to be the Cartan subgroup $U(1)^{N-1}$, for which $C_H=1$,
one finds that for $N\ge 4$ the radical is positive,
and thus the solution~(\ref{sol}) applies,
whereas for $N=2$ and $N=3$, Eq.~(\ref{sol2}) is the relevant
solution.  If we take $H=SU(M)\times SU(N-M)\times U(1)$, the
radical is always negative (with $\g=\sqrt{23}/22$ for all $N$ and $M$).
All solutions are decreasing with increasing $t$,
because $dx/dt$ is always negative according to Eq.~(\ref{ricatti}).
Before we turn to a discussion of each type of solution, we note that in both cases,
if $\tg^2=x-3g^2$ is chosen positive at some scale $t=t_0$, it is guaranteed that $\tg^2$
will stay positive at all lower scales down to the first value of $t$ where the solution
diverges.  This is of course because $d\tg^2/dt$ is always negative.

\ssec{4.1.}{Case $1-4\;\frac{\hb\td}{b^2}>0$}
When the radical is positive, we have that $1>\a_+>\half>\a_->0$, and, with
\begin{equation}
\label{chi}
\chi\equiv\sqrt{1-4\;\frac{\hb\td}{b^2}}<1\ ,
\end{equation}
the solution can be written as ($t>0$)
\begin{equation}
\label{possol}
x(t)=\frac{1}{2\td t}\;\frac{p(1+\chi)t^\chi+q(1-\chi)}{pt^\chi+q}\ .
\end{equation}
Depending on the sign of the integration constant $q/p$, this solution shows
two types of behavior.  When $q/p>0$, this solution is always positive for
$t>0$, and diverges for $t\to 0$.  In other words, the coupling $\tg^2$
diverges at the same scale as the coupling $g^2$.  When $q/p<0$, the solution
already diverges for a positive value of $t$ with $t^\chi=-q/p$, and the
coupling $\tg^2$ diverges at a scale $\tL$ larger than the scale $\L$
at which $g^2$ diverges.

The sign of $q/p$ can be translated into an inequality between the
``initial conditions" $\tg_0^2=\tg^2(t_0)$ and $g_0^2=g^2(t_0)$
where $t_0>0$ determines the ultra-violet cutoff
$\L_0\equiv\L\,{\rm exp}(16\pi^2t_0)$.  In terms of
$x_0=x(t_0)$,  we have
\begin{equation}
\label{pqratio}
-\frac{q}{p}=\frac{t_0^\chi(\td x_0t_0-\a_+)}{\td x_0t_0-\a_-}\ .
\end{equation}
Requiring that $-q/p$ be positive therefore translates into (recall $\a_-<\a_+$)
\begin{equation}
\label{xineq}
x_0>\frac{\a_+}{\td t_0}=\frac{\a_+b}{\td}\;g_0^2\ ,\ \ \ \ \ {\rm or}\ \ \ \ \ x_0<
\frac{\a_-}{\td t_0}=\frac{\a_-b}{\td}\;g_0^2\ .
\end{equation}
As can be seen by combining Eqs.~(\ref{possol}) and~(\ref{pqratio}) into
\begin{equation}
\label{xagain}
x(t)=\frac{1}{\td t}\;\frac{\td x_0 t_0(\a_+t^\chi-\a_-t_0^\chi)-\a_+\a_-(t^\chi-t_0^\chi)}
{\td x_0 t_0(t^\chi-t_0^\chi)-(\a_-t^\chi-\a_+t_0^\chi)}\ ,
\end{equation}
the second of inequalities~(\ref{xineq}) leads to a divergence of $x(t)$
at some $t>t_0$; hence the only divergence at $t<t_0$ occurs at
$t=0$, as for $q/p>0$.  The first of inequalities~(\ref{xineq})
leads to a divergence of $x(t)$
at $t=\tit$ with $0<\tit<t_0$, and thus represents the new class of
solutions with an infra-red scale $\tL=\L\,\exp(16\p^2\tit)$
such that $\L<\tL<\L_0$.
We note that the first of inequalities~(\ref{xineq})
implies that certainly $x_0\ge 3g_0^2$, as required by
Eq.~(\ref{x}), because $\a_+>\half$, and thus $\a_+b/\td>22N/(3N+6C_H)\ge 11/3>3$,
using $0< C_H\le N/2$.

Summarizing, we find solutions diverging at a value of $t$ between zero
and $t_0$ if and only if
\begin{equation}
\label{couplingineq}
\tg_0^2>\frac{1}{2\td}\left(2b\a_+-\tc\right)g_0^2\ ,
\end{equation}
and solutions only diverging at $t=0$ on the interval
$0\le t\le t_0$ if this inequality is not satisfied.
One can verify that $\frac{1}{2\td}\left(2b\a_+-\tc\right)g_0^2\ge \frac{2}{3}g_0^2$.

\ssec{4.2.}{Case $1-4\;\frac{\hb\td}{b^2}<0$}
When the radical is negative,
the solutions behave very differently from those discussed above.
The solutions have an infinite number of zeros,
\begin{equation}
\label{zeros}
x(t'_n)=0\ \ \ \ \ {\rm iff}\ \ \ \ \ \g\log(t'_n)=\phi+\b+n\pi\ ,
\end{equation}
where $\b=\arctan(-2\g)$, and an infinite number of
divergences,
\begin{equation}
\label{infties}
x(t''_n)=\infty\ \ \ \ \ {\rm iff}\ \ \ \ \ \g\log(t''_n)=\phi+n\pi\ ,
\end{equation}
in addition to the divergence at $t=0$.
Clearly, the zeros and divergences alternate, consistent with the fact
that $x$ is always decreasing.  In terms of the ``initial condition"
$x_0=x(t_0)$, $\phi$ can be expressed
as
\begin{equation}
\label{phi}
\phi=\g\log(t_0)-\arctan{\left(\frac{\g}{\td t_0 x_0-\half}\right)}\ .
\end{equation}
The solution $x(t)$ can be written in terms of $x_0$ and $t_0$ by combining Eqs.~(\ref{sol2}) and~(\ref{phi}):
\begin{equation}
\label{xincond}
x(t)=\frac{1}{\td t}\left(\half+\g\;\frac{\td x_0 t_0-\half-\g\tan{(\g\log{(t/t_0))}}}
{(\td x_0 t_0-\half)\tan{(\g\log{(t/t_0))}}+\g}\right)\ .
\end{equation}
We may now define the infra-red scale $\tL$ by choosing $\tit$ as
the first value below $t_0$ where $x(\tit)$ diverges.  It follows
that always $\tL>\L$  ($\tit$ is always larger than zero, because
arbitrarily small positive solutions of
Eq.~(\ref{infties}) can be obtained by taking $n$ large and
negative).

\vspace{5ex}
\noindent {\large\bf 5.~Discussion of implications}
\secteq{5}
\vspace{3ex}

In this section, we discuss the possible implications of the one-loop RG
analysis of the previous two sections.  We emphasize that this discussion is
speculative, because  a one-loop calculation
cannot really tell us what the actual non-perturbative dynamics of the theory looks
like.

The first observation is of course that both $g^2$ and $\tg^2$ are asymptotically
free.  If we choose small values for both couplings at some high scale $t_0$,
they will grow as one goes down in energy.   The dynamics
will be non-perturbative at low scales, while perturbation theory should
give a reliable description of the dynamics at scales well above the larger
of $\L$ and $\tL$.  Clearly, in
order to find out more, non-perturbative techniques, such as numerical lattice
computations, will be needed.   However, we may contemplate what the
resummed one-loop calculation, embodied in the solution to the RG equations,
suggests for the dynamics governed by the coupling $\tg^2$.

One important question, which we already raised in the Introduction, is whether
SSB of eBRST symmetry might take place due to the fact that $\tg^2$
becomes strong.   For example, it is not excluded that some ghost condensate is generated,
reminiscent of the way a chiral condensate is generated in QCD.
A  question is then at what scale this SSB, or any other strong dynamics, would take place.

The results of our one-loop RG analysis give some interesting hints as to
what to expect.  Let us assume that the message
of this analysis is the following: The scale at which a running
one-loop coupling diverges provides a valid estimate of the scale where that
coupling becomes
strong.   The RG analysis thus implies that
$g^2$ cannot become strong without $\tg^2$ becoming strong too, since
the scale $\tL$ associated with $\tg^2$ is always larger than or
(approximately) equal to
$\L$, the scale associated with $g^2$.

First, consider  the simpler case analyzed in Sec.~4.1.  There are
two possibilities: $\tL=\L$ if $\tg_0^2$ is chosen small enough relative
to $g_0^2$;  otherwise, $\tL>\L$
({\it cf.} Eq.~(\ref{couplingineq})).  There is no
upper limit to $\tL$ (as long as it is below the cutoff scale $\L_0$
which, in turn, can be arbitrarily large).
Of course, the precise distinction between these two possibilities results from our
definition of the infra-red scales as those where the one-loop couplings become infinite.
Had we defined the infra-red scales by requiring the running
couplings to take some finite value of order one,
we would still find solutions with
$\tL\sim\L$, and solutions with $\tL\gg\L$, but no solutions with $\tL\ll\L$.
In other words, it is likely that there is a region in the phase
diagram where the couplings become strong simultaneously, and a
region where $\tg^2$ becomes strong while $g^2$ is weak,
with the possibility of choosing $\tL/\L$ arbitrarily large.

For the case analyzed in Sec.~4.2  the situation looks rather strange at first.
If we demand that $\tg^2(t)$ assumes some fixed finite value for some fixed
$t>0$, the cutoff $t_0$ cannot be taken arbitrarily large; in fact
$t_0$ must be kept below the first value of $t$ where the solution vanishes.
In practice, however, the physical interval between one value of $t$
where the solution diverges, and the closest higher value where the
solution vanishes can be very large. The reason is the logarithmic dependence
on $t$, which represents double-logarithmic behavior in scale.  Two
numerical examples (taking $G/H=SU(2)/U(1)$) will suffice to illustrate this.  In both examples
we take $\L=100$~MeV, and $t_0=0.29$, corresponding to
$\L_0\simeq 10^{19}$~GeV.

In the first example, we choose $x_0=1/(2\td t_0)=0.86$ or $\tg_0^2=0.16$, and it follows from Eq.~(\ref{xincond}) that
\begin{equation}
\label{ex1}
x(t)=\frac{1}{2\td t}\left(1-2\g\tan(\g\log(t/t_0))\right)\ .
\end{equation}
The first divergence below $t_0$ happens when $\g\log(t/t_0)=-\pi/2$,
which translates into a value for $\tL$ about 3\% larger than $\L$.
So, we can easily arrange that $\tg^2$ becomes strong at about the same scale as $g^2$,
while choosing the cutoff at the Planck scale.

In the second example, we raise $x_0$ to 1.7 or $\tg_0^2=1.0$, which is probably still a
relatively weak coupling.  This leads to $\tL= 75$~GeV.
We see that, given a cutoff scale of $10^{19}$~GeV,
and choosing a typical QCD scale for $\L$, we can arrange
for $\tL$ to be virtually any scale in between.
In practice, the solutions of Sec.~4.2 are not so different
from the mathematically simpler solutions of Sec.~4.1.

\vspace{5ex}
\noindent {\large\bf 6.~Conclusion}
\secteq{6}
\vspace{3ex}

We calculated the one-loop beta function for the longitudinal coupling $\tg^2=\x g^2$ in an
equivariantly gauge-fixed $SU(N)$ YM theory in which only the Cartan subgroup
$U(1)^{N-1}$ or the subgroup $H=SU(M)\times SU(N-M)\times
U(1)$ is left unfixed.  We found that also
the longitudinal coupling $\tg^2$ is
asymptotically free, just like the gauge coupling $g^2$.
These results readily generalize to other subgroups, {\it cf.}~Appendix~A.

We then solved the corresponding one-loop RG equations.  Since both
couplings are asymptotically free, dimensional transmutation takes place
for both couplings, and an infra-red scale is dynamically generated for each.   In the case
of the YM coupling $g^2$ we know that the one-loop beta function correctly
``predicts" that the coupling indeed becomes strong at low energies.
Our one-loop
results suggest that also the longitudinal coupling $\tg^2$ becomes strong at low energy
when, or even before, the gauge coupling $g^2$ becomes strong.
{\it A priori}, a
possibility might have been that $\tg^2$ stays weak also when $g^2$ becomes
strong, but this possibility seems excluded by the one-loop RG analysis.

The one-loop RG analysis suggests that there is always some
strong-coupling dynamics
(possibly accompanied by SSB and/or the formation of bound states) in the
longitudinal sector.  As already discussed in the Introduction, a key question
is whether there are circumstances in which the strong dynamics
at the scale $\tL$ affects the gauge-invariant sector.
If so, a further question would be how
the modified dynamics in the gauge-invariant sector depends on the scale $\tL$;
\ie\ whether there is any qualitative distinction between the cases $\tL\sim\L$ and $\tL\gg\L$.
Do these two possibilities correspond to different phases?  This is, to us, a question
without an obvious answer.  In the case that $\tL\gg\L$, $\tg^2$ becomes strong
while $g^2$ stays weak, and the strong-coupling dynamics is thus entirely
determined by that of the longitudinal sector (the ``reduced" model, {\it cf.} Sec.~1).
In the case that $\tL\sim\L$, all degrees of freedom may play an important role,
with possibly different implications.

Clearly, the full dynamics of this gauge-fixed theory, including that of the longitudinal sector, will
have to be further investigated by non-perturbative methods.
In addition to providing the first hints of what the dynamics of equivariantly
gauge-fixed YM theories might look like, the one-loop
beta-function analysis presented here provides a useful first step in that it predicts
the scaling behavior in the limit in which the cutoff is taken large.

\vspace{3ex}
\noindent {\bf Acknowledgements.}
MG would like to thank Massimo Testa for some useful discussions.
MG would also like to thank the Department of Physics
of Tel Aviv University, and YS that of San Francisco State University,
for hospitality.
MG is supported in part by the US Department of Energy,
and YS is supported by the Israel Science Foundation under grants
222/02 and 173/05.

\vspace{5ex}
\noindent {\large\bf Appendix A}
\secteq{A}
\vspace{3ex}

\tolerance=2000
Here we prove Eq.~(\ref{H}) and compute $H_\a$ for a large class of subgroups
\mbox{$H\supset U(1)^{N-1}$},
still taking $G=SU(N)$. We will assume that the generators of $H$ fit into
$K$ square blocks along the main diagonal. The size of the $r$-th block is
$M_r\times M_r$, where $1 \le M_r < N$. The number of blocks is constrained
by $2\le K \le N$. In the special case $K=N$ all blocks are $1\times 1$, which
corresponds to $H=U(1)^{N-1}$ \cite{gsnonab}.
In this appendix we thus assume $K<N$,
implying that there is at least one block with size $M_r>1$,
corresponding to a subgroup $SU(M_r)$. The special case $K=2$ corresponds
to the maximal subgroup $SU(M)\times SU(N-M)\times U(1)$
discussed in this paper. The case $K=1$ is excluded as it corresponds to $H=SU(N)=G$.
Note that the number of Cartan generators of $SU(N)$ contained in the non-abelian
subgroups (the blocks with $M_r>1$) is $\sum_{r=1}^K (M_r-1)$.
It will be convenient to assume that the blocks are ordered by increasing
size: $M_1 \le M_2 \cdots \le M_K$, implying in particular
that any $1\times 1$ blocks occur before all the $M_r>1$ blocks.

\tolerance=1000
We next construct a convenient basis for the off-diagonal $SU(N)$ generators,
which includes all coset generators. We define $N\times N$ matrices
$\t_{AB}^k$ where $1 \le A < B \le N$ and $1 \le k \le 3$. The matrix
$\t_{AB}^k$ has precisely two non-zero entries, each with absolute value
equal to $1/2$, chosen such that $\t_{AB}^k$ reduces to (half)
the corresponding Pauli matrix, $\s^k/2$,
if we keep only the $A$-th and $B$-th rows and columns.
The coset generators are spanned by $\t_{AB}^s$, where $s=1,2$,
and where the non-zero entries are constrained to be outside
of all the diagonal blocks. For the maximal subgroups
(assuming $M\le N/2$ to conform to the above ordering),
this implies $1 \le A \le M$ and $M+1 \le B \le N$, and it is easy
to check by inspection that $f_{\a\b\g}=0$.  In this case, one thus
finds $G_\a=C_G=0$.

It is convenient
to make a unitary transformation to the non-hermitian
$\t_{AB}^\pm = (\t_{AB}^1 \pm i\t_{AB}^2)/\sqrt{2}
= (\t_{AB}^\mp)^\dagger$, which satisfy the orthogonality relation
$\tr(\t_{AB}^\mp\, \t_{CD}^\pm) = \half\d_{AC}\d_{BD}$.
The matrices $\t_{AB}^+$ each have a single non-zero entry
located above the main diagonal.
Using this basis, Eq.~(\ref{H}) may be re-written as
\begin{equation}
  -2 \sum_i \tr\left([T^i,\t^\pm_{AB}][T^i, \t^\mp_{CD}]\right)
  = H_{AB}\; \d_{AC}\d_{BD}\,.
\label{HAB}
\end{equation}
Denote the number of $1\times 1$ blocks by $L$.
(Our assumptions imply $0\le L \le N-2$.)
Eq.~(\ref{HAB}) is proved by considering the following three cases:
\begin{itemize}
\item $1\le A<B\le L$ (this requires $L\ge 2$). This effectively
reduces to the case of $H=U(1)^{N-1}$.  Constancy of $H_{AB}$
follows from skewed permutation symmetry \cite{gsnonab}.
The actual numerical value is found to be $H_{AB}=1$.
\item $1\le A\le L < B$. We say that $B$ ``belongs'' to the $r$-th
block if the $B$-th entry along the main diagonal resides in
the $r$-th block. By assumption, the size of this block is $M_r>1$.
It is easy to see that $\t_{AB}^+$ transforms as the anti-fundamental
representation under $SU(M_r)$. We calculate $H_{AB}$ by
summing the contribution from all the Cartan generators of $SU(N)$,
which is always equal to one, plus the contribution of the extra,
off-diagonal $SU(M_r)$ generators. We use the known value of the quadratic
Casimir of the fundamental representation, as well as
$\sum_i T_i^2=(N-1)/(2N)$,
which is true for the sum over the Cartan generators of any $SU(N)$
in the fundamental representation.
This proves Eq.~(\ref{HAB}), with $H_{AB}=1+D(M_r)$ where $D(M)=(M-1)/2$.
\item $L < A < B$. Now $A$ and $B$ belong to the $r$-th and $s$-th
blocks respectively, where $1<M_r\le M_s$.
The matrix  $\t_{AB}^+$ transforms as the fundamental
representation of $SU(M_r)$, and as the anti-fundamental
representation under $SU(M_s)$, and
Eq.~(\ref{HAB}) follows with $H_{AB}=1+D(M_r)+D(M_s)$.
\end{itemize}
As a check, consider $H=SU(M)\times SU(N-M) \times U(1)$.
For $M\ge 2$, this corresponds to the third case, and
$H_{AB}=1+D(M)+D(N-M)=N/2$. For $M=1$ one has
$H=SU(N-1) \times U(1)$, which corresponds to the second case,
and again $H_{AB}=1+D(N-1)=N/2$.  Another example
of a subgroup with a constant $H_{AB}$,
which is a generalization of the case $H=U(1)^{N-1}$, is
$H=SU(M)^K\times U(1)^{K-1}$ with $G=SU(N=MK)$.  In this case,
the subgroup's generators fit into $K$ blocks of size $M\times M$, and
$H_{AB}=M$.


\vspace{1ex}


\begin{thebibliography}{99}

\bibitem{gsnonab}
  M.~Golterman and Y.~Shamir,
  Phys.\ Rev.\ D {\bf 70}, 094506 (2004)
  [arXiv:hep-lat/0404011].

\bibitem{schaden}
M.~Schaden,
Phys.\ Rev.\ D {\bf 59}, 014508 (1999)
[arXiv:hep-lat/9805020].

\bibi{hn}
H.~Neuberger,
Phys.\ Lett.\ B {\bf 183}, 337 (1987).

\bibitem{gfreview}
  L.~Giusti, M.~L.~Paciello, C.~Parrinello, S.~Petrarca and B.~Taglienti,
  Int.\ J.\ Mod.\ Phys.\ A {\bf 16}, 3487 (2001)
  [arXiv:hep-lat/0104012].

\bibi{testa}
M.~Testa,
Phys.\ Lett.\ B {\bf 429}, 349 (1998)
[arXiv:hep-lat/9803025].

\bibi{mg2000}
For the most recent comprehensive review see: M.~Golterman,
Nucl.\ Phys.\ Proc.\ Suppl.\  {\bf 94}, 189 (2001)
[arXiv:hep-lat/0011027].

\bibi{rome}
A.~Borrelli, L.~Maiani, R.~Sisto, G.~C.~Rossi and M.~Testa,
Nucl.\ Phys.\ B {\bf 333}, 335 (1990);
G.~C.~Rossi, R.~Sarno and R.~Sisto,
Nucl.\ Phys.\ B {\bf 398}, 101 (1993).

\bibi{u1}
Y.~Shamir,
Phys.\ Rev.\ D {\bf 57}, 132 (1998)
[arXiv:hep-lat/9512019];
M.~Golterman and Y.~Shamir,
Phys.\ Lett.\ B {\bf 399}, 148 (1997)
[arXiv:hep-lat/9608116];
W.~Bock, M.~Golterman and Y.~Shamir,
Phys.\ Rev.\ Lett.\  {\bf 80}, 3444 (1998)
[arXiv:hep-lat/9709154].

\bibitem{gz}
  M.~Golterman and L.~Zimmerman,
  Phys.\ Rev.\ D {\bf 71}, 117502 (2005)
  [arXiv:hep-lat/0504023].

\bibitem{schaden2}
  M.~Schaden,
  arXiv:hep-th/9909011.

\bibitem{gr}
See for instance
I.S.~Gradshteyn and I.M.~Ryzhik, Table of Integrals, Series, and Products,
Academic Press.

\end{thebibliography}
\end{document}